\documentclass[letterpaper, 11pt]{article}

\usepackage[nocompress]{cite}
\usepackage{jheppub}

\usepackage{graphicx}
\usepackage{epstopdf}
\usepackage{amsmath, amssymb}

\usepackage{bm}
\usepackage{caption}
\usepackage{subcaption}

\usepackage{multirow} 
\usepackage{longtable} 

\usepackage{romannum}
\usepackage{url}

\usepackage[us,12hr]{datetime} 

\newcommand{\be}{\begin{eqnarray}}
\newcommand{\ee}{\end{eqnarray}}

\newcommand{\bn}{\begin{enumerate}}
\newcommand{\en}{\end{enumerate}}

\def\CI{{\cal I}}

\def\CM{{\cal M}}
\def\CN{{\cal N}}
\def\CO{{\cal O}}

\def\CS{{\cal S}}

\def\k{\kappa}

\def\s{\sigma}

\def\G{\Gamma}

\def\half{\frac{1}{2}}

\def\Tr{{\rm Tr}}
\def\tr{{\rm Tr}}
\def\det{{\rm det}}

\def\PE{\textrm{PE}}

\def\vec#1{\bm{#1}}

\newcommand{\bea}{\begin{eqnarray}}
\newcommand{\eea}{\end{eqnarray}}

\def\CI{{\cal I}}

\def\CM{{\cal M}}
\def\CN{{\cal N}}
\def\CO{{\cal O}}

\def\CS{{\cal S}}

\def\k{\kappa}

\def\s{\sigma}

\def\G{\Gamma}

\def\half{\frac{1}{2}}

\def\ft{\mathfrak{t}}

\def\Tr{{\rm Tr}}
\def\tr{{\rm Tr}}
\def\det{{\rm det}}
\def\PE{{\rm PE}}

\allowdisplaybreaks

\title{A ``Lagrangian" for the $E_7$ Superconformal Theory}

\author[a]{Prarit Agarwal,}
\author[b]{Kazunobu Maruyoshi,}
\author[c, d]{and Jaewon Song}
\affiliation[a]{Department of Physics and Astronomy \& Center for Theoretical Physics\\ Seoul National University, Seoul 08826, Korea}
\affiliation[b]{Faculty of Science and Technology, Seikei Unversity\\ 3-3-1 Kichijoji-Kitamachi, Musashino-shi, Tokyo, 180-8633, Japan}
\affiliation[c]{School of Physics, Korea Institute for Advanced Study\\
85 Hoegiro, Dongdaemun-gu, Seoul 02455, Korea}
\affiliation[d]{Department of Physics, University of California, San Diego\\
9500 Gilman Dr, La Jolla, CA 92093, USA}
\emailAdd{agarwalprarit@gmail.com}
\emailAdd{maruyoshi@st.seikei.ac.jp}
\emailAdd{jsong@kias.re.kr}

\abstract
{
We find an $\CN=1$ gauge theory that flows to the rank-one $\CN=2$ superconformal field theory with $E_7$ flavor symmetry. We first obtain a Lagrangian description for the $R_{0, N}$ theory, which appears in the S-dual description of the $SU(N)$ gauge theory with $2N$ fundamental hypermultiplets. This is a straightforward generalization of the proposed Lagrangian description for the $E_6$ theory. The $E_7$ theory is then obtained via partial Higgsing of the $R_{0, 4}$ theory. From this Lagrangian description, we compute the full superconformal index. We also consider twisted dimensional reduction on $S^2$ to obtain $\CN=(0, 4)$ theory for the $E_7$ one instanton string and compute its elliptic genus. 
 
}

\preprint{SNUTP18-001, KIAS-P18021}

\begin{document}
\maketitle

\section{Introduction}

In recent years, it has become evident that there exists a vast landscape of non-conventional strongly-coupled theories that are not readily describable in terms of conventional Lagrangian path-integral approach. These so-called ``non-Lagrangian" theories do not admit a weakly coupled description that can be studied using perturbation theory. These theories are mostly obtained via string/M-theory construction or realized as a special limit of certain known Lagrangian field theories. Nevertheless, plenty of quantitative properties have been obtained via various methods mostly thanks to string/M-theory origin of the theory. 

More recently, it was discovered that certain non-Lagrangian 4d $\CN=2$ SCFTs can be actually obtained as infrared theories of $\CN=1$ gauge theories, where supersymmetry gets enhanced at the fixed points. This microscopic realization provides a useful direct way of investigating quantitative nature of such theories. Two distinct classes of examples have been found. One is the case of $E_6$ SCFT, that is obtained in a way that utilizes S-duality in a crucial way \cite{Gadde:2015xta}. This gave a physical interpretation of the Spiridonov-Warnaar inversion formula of the elliptic beta integral \cite{SPIRIDONOV200691}. The other is the Argyres-Douglas (AD) SCFTs, which are realized as fixed points of adjoint SQCDs or quiver gauge theories with a number of gauge singlets, deformed by dangerously irrelevant operators \cite{Maruyoshi:2016tqk, Maruyoshi:2016aim, Agarwal:2016pjo,Benvenuti:2017lle,Agarwal:2017roi,Benvenuti:2017kud,Benvenuti:2017bpg}\footnote{Also see \cite{Aghaei:2017xqe}, where it has been shown, following the integral identities in \cite{deBult:2007}, that upon dimensional reduction, the $S^3$ partition function of the these Lagrangians matches exactly with that of the 3d mirror of the corresponding AD theory. }.

In this paper, we 
generalize the result of \cite{Gadde:2015xta} to the $SU(N)$ case to obtain a Lagrangian description for the $R_{0, N}$ theory \cite{Chacaltana:2010ks}, which reduces to the Minahan-Nemechansky's $E_6$ theory \cite{Minahan:1996fg} for $N=3$. Then by considering a partial Higgsing for the $R_{0, 4}$ theory, we obtain the $E_7$ theory \cite{Minahan:1996cj}. 
These ``Lagrangians" enable us to compute the supersymmetric partition functions of the $R_{0,N}$ and the $E_7$ theories. 
We compute the superconformal index of the $E_7$ theory, which can be considered as a straightforward generalization of \cite{Gadde:2010te} for the $E_6$ theory. 

We also consider a twisted dimensional reduction of the $R_{0, N}$ theory and the $E_7$ theory to 2d $\CN=(0, 4)$ SCFT \cite{Putrov:2015jpa, Gadde:2015wta}. Especially the $E_7$ theory describes the one instanton string of $E_7$ group, from which we compute its elliptic genus and compare with other results in existing literature. 

This paper is organized as follows: We describe the matter content and interactions for the $R_{0, N}$ and $E_7$ theory in section \ref{sec:Lag}. Following this, we compute the full superconformal indices for  these theories in section \ref{sec:Index} and compare with the known results in special limits. In section \ref{sec:2d}, we consider twisted dimensional reduction to 2d and compute the elliptic genus of the $E_7$ theory and compare with the known result. We then conclude with some remarks. Some detailed results of the computations are given in appendices.  


\section{Lagrangian for the $R_{0, N}$ and $E_7$ theory}
\label{sec:Lag}

\subsection{$R_{0, N}$ theory} 
\label{subsec:R0NLag}
  The $R_{0,N}$ theory is a 4d $\CN=2$ SCFT with flavor symmetry $SU(2N) \times SU(2)$ and central charges 
    \bea
    a
     =     \frac{7 N^2 - 22}{24}, ~~~
    c
     =     \frac{2 N^2 - 5}{6}.
     \label{acR0N}
    \eea
  The chiral ring is generated by the following generators:
  \begin{itemize}
  \item Moment map operators, $\mu_{SU(2N)}$ and $\mu_{SU(2)}$, 
           transforming in the adjoint representations of $SU(2N)$ and $SU(2)$ respectively.
           The $(I_3, r)$ charge for both of them is given by  $(1, 0)$. Thus they both have a  scaling dimension given by $\Delta(\mu_{SU(2N)}) = \Delta(\mu_{SU(2)})=2$.
  \item $Q_{(N)}$, transforming in the $(\wedge^N, \bar{\mathbf{2}})$ representation of $SU(2N) \times SU(2)$,
           with $(I_3, r)=(\frac{N-1}{2}, 0)$ and $\Delta(Q_{(N)})=N-1$.
           These three parametrize the Higgs branch of the $R_{0,N}$ theory.
  \item The Coulomb branch operators, $u_d$, where $d=3,4,\ldots,N$,
           with $(I_3, r) = (0, 2d)$ and $\Delta= d$.
  \end{itemize}
 Here $\wedge^k$ is the $k$-index anti-symmmetric tensor representation
  and $I_3$ and $r$ are the charges of the Cartan of $SU(2)_R$ and $U(1)_r$ of $\CN=2$ $R$-symmetry respectively.
  
  The Coulomb branch is freely generated.
  On the other hand, the Higgs branch operators satisfy various chiral ring relation.
  For operators with low scaling dimensions, these relations are as follows:
    \bea
    (\mu_{SU(2N)}^2)^f_{~f'}
    &=&     \tr \mu_{SU(2)}^2 \delta^f_{f'},
     \label{relation1R} \\
    (\mu_{SU(2N)})^{f_1}_{~f'} Q_{(N)~~~~~~~\alpha}^{f',f_2,\ldots, f_N}
    &=&    \frac{1}{N} (\mu_{SU(2)})^\beta_{~\alpha} Q^{f_1,f_2,\ldots,f_N}_{(N)~~~~~~~\beta},
               \label{relation2R} \\           
    (\mu_{SU(2N)})^{f'}_{~f_1} \epsilon_{f',f_2,\ldots,f_{2N}} Q_{(N)~~~~~~~~~~~~~~\alpha}^{f_{N+1},f_{N+2},\ldots, f_{2N}}
    &=&    \frac{1}{N} (\mu_{SU(2)})^\beta_{~\alpha} \epsilon_{f_1,\ldots,f_{2N}} Q^{f_{N+1},\ldots,f_{2N}}_{(N)~~~~~~~\beta},
                \label{relation3R} \\
    X u_d
    &=&    0,
    \eea
  where $X$ is arbitrary Higgs branch operator. 
  The chiral ring relations were studied also in \cite{Collinucci:2017bwv,Hanany:2017pdx}.

\paragraph{S-duality}
  The $R_{0,N}$ theory was originally found as the strong coupling limit of the $\CN=2$ $SU(N)$ SQCD 
  with $2N$ fundamental hypermultiplets \cite{Chacaltana:2010ks}, by generalizing the arguments in \cite{Argyres:2007cn,Gaiotto:2009we}.
  More precisely, the $\CN=2$ $SU(N)$ gauge theory with $2N$ fundamental hypermultiplets is dual to the theory obtained by gauging the $SU(2)\subset SU(2) \times SU(2N)$ flavor symmetry of $R_{0, N}$ and coupling it to   a single $SU(2)$ fundamental hypermultiplet \cite{Chacaltana:2010ks}. 
  When $N=3$, it reduces to the Argyres-Seiberg duality \cite{Argyres:2007cn} between $SU(3)$ $N_f=6$ theory and the $E_6$ SCFT. 
  In Appendix \ref{subsec:duality}, we will show that the chiral ring generators and the relations are consistent with the duality.
   
  From the class $\CS$ perspective \cite{Gaiotto:2009hg,Gaiotto:2009we}, the $R_{0,N}$ theory is realized as a low energy effective theory obtained 
  by compactifying the 6d $\CN=(2,0)$ theory of type $A_{N-1}$ on a three-punctured Riemann sphere with two maximal punctures
  and the third puncture being specified by the L-shaped partition $[N-2, 1^2]$.
  The flavor symmetry carried by the punctures are $SU(N)^2 \times U(1) \times SU(2)$, 
  which is indeed a maximal subgroup of the full flavor symmetry $SU(2N) \times SU(2)$.

\paragraph{Lagrangian for $R_{0,N}$}
  We now describe the $\CN=1$ Lagrangian which flows to the $R_{0,N}$ theory in the infrared.
 To do so, we apply the deformation described in \cite{Gadde:2015xta} verbatim 
  to the $\mathcal{N}=2$ $SU(N)$ gauge theory with $2N$ fundamental hypermultiplets. 
  This deformation procedure utilizes the S-duality described above.
  Indeed in order to get the $R_{0,N}$ theory on the dual side of SQCD, what we have to do is to ungauge the $SU(2)_g$ gauge part
  and decouple the additional hypermultiplet denoted here by ($\mathfrak{q},\tilde{\mathfrak{q}}$).
  This is easily done in the following manner:
  we first turn off the coupling for the superpotential term $\mathfrak{q} \phi_D \tilde{\mathfrak{q}}$, 
  where $\phi_D$ is the $\mathcal{N}=1$ chiral multiplet residing in the $\mathcal{N}=2$ vector multiplet for $SU(2)_g$. 
  Upon doing this, the $U(1)_s$ flavor symmetry acting on the hypermultiplets is enhanced to $SU(2)_s$.
  We further gauge this $SU(2)_s$ by an $\CN=1$ vector multiplet and add two chiral doublets of $SU(2)_s$.
  This step will bestow us with an additional $SU(2)_w$ flavor symmetry acting on the newly added doublets of $SU(2)_s$. 
  We will denote these new $SU(2)_s$ doublets as $(\mathfrak{p},\tilde{\mathfrak{p}})$ with $\mathfrak{p}$ and $\tilde{\mathfrak{p}}$ having $SU(2)_s$-spin $+ \half$  and $-\half$ respectively.
 From the point of view of the $SU(2)_s$ gauge group, we now have an $\mathcal{N}=1$ SQCD with $N_f=N_c=2$. This implies that at low energies, the $SU(2)_s$ gauge group will confine with a quantum deformed moduli space constraint ${\rm Pf} \CM = \Lambda^4$ \cite{Seiberg:1994bz}. Here $\CM$ are the mesons/baryons constructed from $\mathfrak{q}$, $\tilde{\mathfrak{q}}$, $\mathfrak{p}$ and $\tilde{\mathfrak{p}}$
  and $\Lambda$ is the dynamical scale of the $SU(2)_s$ gauge symmetry.
  One can now choose the vacuum where $SU(2)_g \times SU(2)_w$ symmetry is broken down to its diagonal $SU(2)$ group 
  by adding gauge singlet fields $T$ and $T'$ coupled through the superpotential terms $T b+ T' \tilde{b}$, 
  where $b = \mathfrak{p} \tilde{\mathfrak{p}}$ and $\tilde{b} = \mathfrak{q} \tilde{\mathfrak{q}}$ with $SU(2)_s$ indices contracted appropriately. 
  Finally, the remaining $\phi_D$ can be integrated out by coupling it to a new chiral field $\mu$, via a superpotential mass term $\mu \phi_D$.
  By these operations, one recovers the $R_{0,N}$ theory. 
  
  By the deformation, the $U(2)_R$ symmetry is broken to its Cartans $U(1)_r \times U(1)_t$, where $U(1)_t$ is defined by $t = I_3 - \frac{r}{2}$.
  The $U(1)_t$ charge of $\mathfrak{q}$ and $\phi_D$ are $\half$ and $1$ respectively. 
  That of $\mathfrak{p}$ is determined by requiring that the $SU(2)_g \times SU(2)_w \times U(1)_t$ enhances to 
  $SU(4)$ when we turn off the gauge coupling of $SU(2)_g$.
  Thus the $U(1)_t$ charge is $-\half$.
  
  Let us consider this deformation on the SQCD side. 
  We will denote the $SU(N)$ vector multiplet and hypermultiplets in the $\CN=1$ notation as $(W_\alpha$, $\Phi)$ and $(Q$, $\tilde{Q})$ respectively.
  Taking the superpotential coupling to zero in the dual side maps to the large coupling limit of $Q \Phi \tilde{Q}$  term in the SQCD side. 
  In this limit, the enhanced $SU(2)_s$ symmetry appears, 
  which we gauge by an $\CN=1$ vector multiplet $W_\alpha'$ and introduce two copies of the $SU(2)_s$ doublet
  $(\mathfrak{p},\tilde{\mathfrak{p}})$.
  We further add the singlet fields $T$, $T'$ and $\mu$
  and three flipping terms $T B+ T' B' + \mu \phi$, where $B$ and $B'$ are made out of $Q, \tilde{Q}$ and 
  $\mathfrak{p}$, $\tilde{\mathfrak{p}}$ respectively.
  Here $\phi$ is some operator that transforms in the adjoint of the emergent $SU(2)_s$. 
  The various fields appearing in the Lagrangian, along with their charges are given in Table \ref{tab:R4Lag}. 
  Here we simplified the presentations of \cite{Gadde:2015xta} by making the $SU(2N)$ global symmetry manifest. 
  The $U(1)_r$ 
  will mix with $U(1)_t$  to give the $R$-symmetry at the fixed point in the infrared, the exact linear combination being determined by $a$-maximization \cite{Intriligator:2003jj}.

\begin{table} 
	\centering
	\begin{tabular}{|c||c|c||c|c|c|c|}
		\hline
		Field & $SU(N)$& $U(1)_s \subset SU(2)_s$ & $SU(2N)$ & $U(1)_{r}$ & $U(1)_t$ & $SU(2)_w$\\ 
		\hline \hline
		$W_{\alpha}$ & $\rm{adj}$ & 0 &$\mathbf{1}$ & 1 & 0 & $\mathbf{1}$ \\ 			
		$\Phi$ & $\rm{adj}$ & 0 &$\mathbf{1}$ & 2 & $-1$ & $\mathbf{1}$ \\ 
		$Q$ & $\mathbf{N}$ & $\frac{1}{N}$ & $\mathbf{2N}$ & 0& $\half$ & $\mathbf{1}$ \\
		$\widetilde{Q}$& $\mathbf{\bar{N}}$ & $-\frac{1}{N}$ & $\mathbf{\overline{2N}}$ &0& $\half$ & $\mathbf{1}$ \\ 
		\hline 
		$W'_{\alpha}$ & $\mathbf{1}$ & $\{2,0,-2\}$ & $\mathbf{1}$ & 1 & 0 & $\mathbf{1}$ \\ 
		$\mathfrak{p}$ & $\mathbf{1}$ & 1 & $\mathbf{1}$ & 0& $-\half$&  $\mathbf{2}$ \\ 
		$\tilde{\mathfrak{p}}$ & $\mathbf{1}$ & $-1$ &$\mathbf{1}$ & 0& $-\half$ & $\mathbf{2}$ \\
		\hline 
		$T$ & $\mathbf{1}$ &0 &$\mathbf{1}$ & 2& $1$ & $\mathbf{1}$ \\
		$T'$ & $\mathbf{1}$ &0&$\mathbf{1}$ & 2& $-1$ & $\mathbf{1}$ \\
		$\mu$ & $\mathbf{1}$ &0& $\mathbf{1}$ & 0& $1$ & $\mathbf{3}$ \\  
		\hline
	\end{tabular} 
	\caption{Supermultiplets appearing in the Lagrangian for $R_{0, N}$.} 
	\label{tab:R4Lag}   
\end{table}

  Note that the $U(1)_s$ charges of $Q$ and $\tilde{Q}$ are determined by computing the anomalies on the both sides of the original duality as follows:
  since on the dual side, the multiplets $\mathfrak{q}$ and $\tilde{\mathfrak{q}}$ become a doublet of the $SU(2)_s$,
  their $U(1)_s$ charges should be normalized as $\pm 1$. 
  Therefore the anomaly $\tr R_{\CN=1} U(1)_s^2 = (2/3-1)\cdot 2  \cdot 2 = -4/3$.
  On the SQCD side, letting the $U(1)_s$ charges of $Q$ and $\tilde{Q}$ be $\pm s_Q$, the corresponding anomaly is computed to be 
  $\tr R_{\CN=1} U(1)_s^2 = (2/3-1) \cdot s_Q^2 \cdot 2N \cdot N= -\frac{4N^2 s_Q^2}{3}$.
  This fixes the $s_Q= \frac{1}{N}$.


\paragraph{$a$-maximization} Let us parametrize the $R$-charge as $R=r + S t$, where the numerical value of $S$ is to be determined by $a$-maximization. The trial central charges are then given by
\begin{align}
a_{\rm trial} &= -\frac{3}{64} (S-2) \left(N^2 \left(3 S^2+6 S-4\right)-21 S^2+12 S-8\right) \ ,  \\
c_{\rm trial} &= 
-\frac{1}{64} (S-2) \left(N^2 \left(9 S^2+18 S-8\right)-63 S^2+36 S-16\right) \ . 
\end{align}
It is straightforward to check that $a_{\rm trial}$ gets maximized at $S=4/3$. For this value of $S$ we find that the central charges are identical to those of the $R_{0,N}$ theory \eqref{acR0N}.

\subsection{$E_7$ theory from Higgsing of $R_{0, 4}$}

  The $R_{0, 4}$ theory has an $SU(8) \times SU(2)$ flavor symmetry. 
  Upon giving a nilpotent vev to the moment map operator $\mu_{SU(2)}$ of the $SU(2)$ part of the flavor group, the theory flows to $E_7$ SCFT in the infrared. From the class $\CS$ perspective, this is tantamount to changing the shape of Young diagrams for a three-punctured sphere of type $([1^4], [1^4], [2, 1^2])$ to $([1^4], [1^4], [2^2])$. The resulting theory has a manifest flavor symmetry given by $SU(4)\times SU(4) \times SU(2)$ which enhances to $E_7$. 

  We can also consider an analogous nilpotent Higgsing of $R_{0,N}$ theory. 
  Let us call the resulting theory $\tilde{R}_{0, N}$.
  In the class $\CS$ language, it arises from the three-punctured sphere with two maximal punctures and one puncture of type $[N-2, 2]$. 
  It becomes the $E_7$ theory for $N=4$. 
  
  From the perspective of the ``Lagrangian" in the previous subsection, we give a nilpotent vev to the moment map operator $\mu$ for the $SU(2)_w$ flavor symmetry. 
  This can be done by following the same procedure as in \cite{Gadde:2013fma, Agarwal:2014rua}. Before Higgsing, we have the superpotential term $W \supset \mu \phi$. Now, we give a nilpotent vev $\s^+ = \s^1 + i \s^2$ to $\mu$, where $\s^i$ are the Pauli matrices. This will shift the $U(1)_t$ charge by $t \to t - 2w_3$ where $w_3$ denotes the weight of the $SU(2)_w$. Two out of three components of $\mu$ (with $w_3=0, 1$) will be decoupled after Higgsing and we will be only left with the one with $w_3 = -1$, which we denote as $\mu'$.\footnote{See section 3.3 of \cite{Agarwal:2015vla} for a detailed explanation of the nilpotent Higgsing that is identical to the current case.}
This yields the matter content for the $\tilde{R}_{0, N}$ theory as in table \ref{tab:E7Lag}. 
\begin{table} 
	\centering
	\begin{tabular}{|c||c|c||c|c|c|c|}
		\hline
		Field & $SU(N)$& $U(1)_s \subset SU(2)$ & $SU(2N)$ & $U(1)_{r}$ & $U(1)_t$ \\ 
		\hline \hline
		$W_{\alpha}$ & $\rm{adj}$ & 0 &$\mathbf{1}$ & 1 & 0 \\
		 $\Phi$ & $\rm{adj}$ & 0 &$\mathbf{1}$ & 2 & $-1$\\ 
		$Q$ & $\mathbf{N}$ & $\frac{1}{N}$ & $\mathbf{2N}$  & 0& $\half$  \\
		$\widetilde{Q}$& $\mathbf{\bar{N}}$ & $-\frac{1}{N}$ & $\mathbf{\overline{2N}}$ & 0& $\half$ \\ 
		\hline 
		$W'_{\alpha}$ & $\mathbf{1}$ & $\{2,0,-2\}$ &$\mathbf{1}$ & 1 & 0 \\ 
		$\mathfrak{p}, \tilde{\mathfrak{p}}$ & $\mathbf{1}$ & $\{1, -1\}$ & $\mathbf{1}$ & 0& 0\\ 
		$\mathfrak{p}', \tilde{\mathfrak{p}}'$ & $\mathbf{1}$ & $\{1, -1\}$ & $\mathbf{1}$ & 0& $-1$\\ 
		\hline 
		$T$ & $\mathbf{1}$ &0 &$\mathbf{1}$ & $2$& $1$ \\
		$T'$ & $\mathbf{1}$ &0 &$\mathbf{1}$ & $2$& $-1$ \\
		$\mu'$ & $\mathbf{1}$ &0& $\mathbf{1}$ & $0$ & $2$ \\
		\hline
	\end{tabular} 
	\caption{Supermultiplets appearing in the Lagrangian for $\tilde{R}_{0, N}$. It becomes the $E_7$ theory for $N=4$.} 
	\label{tab:E7Lag}   
\end{table}
This matter content is consistent with the formula for the index we discuss in section \ref{sec:Index}. 

The trial central charges are given as 
\begin{align}
 a_{\rm trial} &= -\frac{3}{64} (S-2) \left(N^2 \left(3 S^2+6 S-4\right)-42 S^2+6 S-4\right) \ ,  \\
 c_{\rm trial} &= -\frac{1}{64} (S-2) \left(N^2 \left(9 S^2+18 S-8\right)-2 \left(63 S^2-9 S+4\right)\right) \ . 
\end{align}
Upon maximizing the trial $a$-function, we again obtain $S=\frac{4}{3}$. This gives us the central charges to be
\begin{align}
 a = \frac{7N^2 - 53}{24} \ , \qquad c = \frac{2N^2 - 13}{6} \ . 
\end{align}
When $N=4$ we obtain $(a, c) = (\frac{59}{24}, \frac{19}{6})$. They agree with the corresponding values for the $E_7$ SCFT \cite{Aharony:2007dj, Shapere:2008zf}. 

\section{The full superconformal index} 
\label{sec:Index}

\subsection{Index of the $R_{0, N}$ theory}
The full superconformal index of the ${R}_{0, N}$ theory can be computed using the Lagrangian described in section \ref{subsec:R0NLag}. This is equivalent to starting with the index of 4d $\mathcal{N}=2$ $SU(N)$ gauge theory with $2N$ fundamental hypermultiplets and using the Spiridonov-Warnaar inversion formula \cite{SPIRIDONOV200691} to extract the index of $R_{0, N}$, exactly in the same way as the index for the $E_6$ theory was computed in \cite{Gadde:2010te}. We will go through the details of this computation explicitly in this section.

The $\mathcal{N}=2$ superconformal index is defined as 
\begin{align} \label{eq:indexDef}
\begin{split}
\mathcal{I} &= \Tr (-1)^F p^{j_1 + j_2 + \frac{r}{2}} q^{j_2 - j_1 + \frac{r}{2}} t^{I_3-\frac{r}{2}} = {\rm{Tr}} (-1)^F \ft^{2(E+j_2)} y^{2j_1} v^{(\frac{r}{2}-I_3)}\ ,
\end{split}
\end{align}
  where $I_3$ and $r$ are the charges of the Cartan of the $SU(2)_R$ and $U(1)_r$ of 4d $\mathcal{N}=2$ superconformal algebra respectively,
  while $j_1,j_2$ are the two Cartans of the 4d Lorentz group $SO(3,1) \sim SU(2)_1 \times SU(2)_2$. 
  The fugacities $(p, q, t)$ of the first definition can be related to the second one $(\ft, y, v)$ via $p=\ft^3 y, q=\ft^3/y, t=\ft^4/v$. 
  
The superconformal index of the 4d $\mathcal{N}=2$ $SU(N)$ gauge theory with $2N$ fundamental hypermultiplets can then be written as 
\be
\begin{split}
	\mathcal{I}_{SU(N)}(s, \vec{a})
	 =  \frac{\left(\kappa \Gamma\left(\frac{pq}{t}\right)\right)^{N-1}}{N!} \oint [d\vec{x}] \frac{\displaystyle\prod_{i=1}^N \prod_{f=1}^{2N}  \Gamma \left(t^{\half} \left( s^{\frac{1}{N}} x_i a_f \right)^{\pm 1}\right)  \prod_{i\neq j} \Gamma \left(\frac{pq}{t} \frac{x_i}{x_j}\right) }{\displaystyle\prod_{i\neq j} \Gamma \left(\frac{x_i}{x_j}\right)} \ ,  
\end{split}
\ee 
where $s$ and $\vec{a}$ are the fugacities of $U(1)_s$ and $SU(2N)$ respectively, and $ [d\vec{x}] = \prod_{i=1}^{N-1} \frac{d x_i}{2 \pi i x_i}$, $\kappa = (p; p)(q; q)$. The repeated exponent denotes we multiply terms of each sign $\G( z^{\pm 1}) \equiv \G(z^1)\G(z^{-1})$. The elliptic gamma function is defined as
\begin{align}
 \G(z) \equiv \G(z; p, q) = \prod_{m, n \ge 0} \frac{1-z^{-1} p^{m+1} q^{n+1}}{1-z p^m q^n} \ . 
\end{align}
Each factors of $\Gamma( (pq)^{r/2} t^{T} \ldots)$ denotes a chiral multiplet of $U(1)_r$ charge $r$ and $U(1)_t$ charge $T$. The terms in the denominator and $\k$ comes from the vector multiplet and $1/N!$ comes from the Weyl group of the gauge group $SU(N)$. 

As mentioned earlier, this theory admits an S-dual frame consisting of a single $SU(2)$ fundamental hypermultiplet coupled to the $R_{0, N}$ theory via an $\mathcal{N}=2$ gauging of its $SU(2) \subset SU(2) \times SU(2N)$ flavor symmetry. 
If we now let $\mathcal{I}_{R_{0, N}} (w,\vec{a})$ represent the index of $R_{0, N}$ with $w$ being the fugacity of $SU(2)$, then $\mathcal{I}_{SU(N)}$ can be written as 
\be
\label{eq:Sduality}
\mathcal{I}_{SU(N)}(s, \vec{a}) = \frac{\kappa \Gamma \left(\frac{pq}{t} \right)}{2}\oint \frac{d w}{2 \pi i w} \frac{  \Gamma \left(t^{\half} s^{\pm1} w^{\pm 1}\right) \Gamma \left(\frac{pq}{t} w^{\pm 2}\right) }{ \Gamma \left(w^{\pm 2}\right)} \mathcal{I}_{R_{0, N}} (w , \vec{a} ) \ .
\ee  
As in the case of \cite{Gadde:2010te}, we can use the Spiridonov-Warnaar inversion formula to obtain the index of the $R_{0, N}$ theory. This gives
\begin{align}
\mathcal{I}_{R_{0, N}} (w, \vec{a} ) = \frac{\kappa}{2\Gamma \left(\frac{pq}{t} w^{\pm 2}\right)}  \oint_{C_s} \frac{d s}{2 \pi i s} \frac{  \Gamma \left(t^{-\half} s^{\pm1} w^{\pm 1}\right) }{ \Gamma\left( t^{-1} \right)\Gamma\left(s^{\pm 2}\right)} \mathcal{I}_{SU(N)} (s,\vec{a}) \ , 
\end{align}
where the contour $C_s$ includes the poles at $s= w^{\pm 1} t^{-1/2}$ but not the ones at $s=w^{\pm 1}  t^{1/2}$. 
  We find that this is indeed the index computed from the matter content in the table \ref{tab:R4Lag}.
To see this, note that
\begin{align}
 \frac{1}{\Gamma \left(\frac{pq}{t} w^{\pm 2}\right) \G(t^{-1})} = \G \left( t w^{\pm 2} \right) \G(pq t)  
 =  \G \left( t w^{\pm 2, 0} \right) \G\left(\frac{pq}{t} \right) \G(pq t) \ , 
\end{align}
where we used the fact that $\G(\frac{pq}{z}) \G(z) = 1$. These 3 terms come from the singlet fields $\mu, T, T'$. 

Evaluating the above integral for $N=4$ explicitly up to $\mathcal{O}(\ft^8)$ (after substituting $p=\ft^3 y, q=\ft^3/y, t=\ft^4/v$), we obtain
\begin{align}
\label{eq:R4index}
\begin{split}
	\mathcal{I}_{R_{0, 4}} &=1+\frac{\ft^4 }{v} \left(\chi^{SU(8)}_{\rm adj} + \chi^{SU(2)}_{\rm adj} \right) 
	+\ft^6 \left(-\chi^{SU(8)}_{\rm adj}-1-\chi^{SU(2)}_{\rm adj}+v^3+ \frac{\chi^{SU(8)}_{70}  \chi^{SU(2)}_2 }{v^{3/2}}\right) \\
	&~+\ft^7 \left(-v^2 \chi^{SU(2)}_2 (y) +\frac{\chi^{SU(8)}_{\rm adj}+1 + \chi^{SU(2)}_{\rm adj} }{v} \chi^{SU(2)}_2 (y) \right)\\
	&~+\ft^8 \left(2 v+v^4-\frac{\chi^{SU(8)}_{70} \chi^{SU(2)}_2}{\sqrt{v}}+\frac{\chi^{SU(8)}_{1232}+\chi^{SU(8)}_{720}+1+ \chi^{SU(8)}_{\rm adj}  \chi^{SU(2)}_{\rm adj} + \chi^{SU(2)}_5}{v^2}\right) \\
	&~+ \CO(\ft^9)  \ ,
\end{split}
\end{align}
where the symbol $\chi^{SU(2)}_n$ (with no argument) denotes the character for the $n$-dimensional representation of the flavor $SU(2)$ and the symbol $\chi_n^{SU(2)} (y)$ with argument $y$ denotes the character for the $n$-dimensional representation $SU(2)_1$ part of the Lorentz group. A similar notation has been used to denote the characters for various representations of the $SU(8)$ flavor symmetry of $R_{0,4}$. 
We present the index for the $R_{0, 4}$ theory up to order $\mathcal{O}(\ft^{18})$ in appendix \ref{app:IR4}. 
  These are consistent with the chiral ring stated in the beginning of section \ref{subsec:R0NLag}.

\subsection{Index of the $E_7$ theory}
To obtain the index of the $E_7$ theory from that of $R_{0,4}$, first recall that for $T_N$ theories the superconformal index is expected to be of the form \cite{Gadde:2009kb,Gadde:2011ik,Gadde:2011uv,Gaiotto:2012xa}
\be
\mathcal{I}_{T_N} (\vec{a},\vec{b}, \vec{c}) = \sum_{\lambda} C_{\lambda} \psi_{\lambda}(\vec{a}) \psi_{\lambda}(\vec{b}) \psi_{\lambda}(\vec{c}),
\ee   
where $\vec{a},\vec{b}$ and $\vec{c}$ are the fugacities of the respective $SU(N)$ group associated to the three punctures of the $T_N$ theory and $\psi_{\lambda}(\vec{a})$ are symmetric functions of the $N$ variables $\vec{a}=(a_1, \cdots, a_N)$ with $\lambda$ being the label for irreducible representations of $SU(N)$.  It is useful to rewrite $\psi_{\lambda}(\vec{a})$ as 
\be
\psi_{\lambda}(\vec{a}) = K(\vec{a}) \Psi_{\lambda}(\vec{a}) .
\ee 
If we now Higgs the puncture according to a partition $\Lambda$, then the index of the theory with the reduced puncture can be obtained from that of $\mathcal{I}_{T_N}$ by $\Psi(\vec{a}) \rightarrow \Psi(\vec{u} t^\Lambda)$ and $K(\vec{a})\rightarrow K_{\Lambda}(\vec{u})$, where we follow the notation of \cite{Gadde:2013fma}. $K_{\Lambda}(\vec{u})$ in the Macdonald limit was obtained in \cite{Gadde:2011uv, Mekareeya:2012tn} and the general form was conjectured in \cite{Gadde:2013fma} to be
\be
K(\vec{a}) \rightarrow
K_\Lambda (\vec{u}) = \PE \left[ \sum_j \frac{t^{1+j}-p q t^{j}}{(1-p)(1-q)} \chi_{R_j} (\vec{u}) \right] .
\ee 
Here, $R_j$'s are irreducible representations of the flavor symmetry that arise upon decomposing the adjoint representation of the $SU(N)$ according to the $SU(2)$ embedding specified by the partition $\Lambda$:  
\begin{align}
 \textrm{adj} \to \bigoplus_j R_j \otimes V_j ,
\end{align}
where $V_j$ are the $SU(2)$ spin-$j$ irreducible representations.
This expression is identical to the contribution of the conserved current multiplet. 

By using this and the fact that both the $R_{0, 4}$ theory and the $E_7$ theory can be obtained from $T_4$, we expect 
\be
\mathcal{I}_{R_{0, 4}} = K_\Lambda (b, w) \sum_{\lambda } C_{\lambda} \Psi_{\lambda}( (b, w) t^\Lambda ) \psi_{\lambda}(\vec{b}) \psi_{\lambda}(\vec{c}) \ , \quad  \Lambda = [2,1,1] \ ,
\ee 
and 
\be
\mathcal{I}_{E_7} = K_{ \Lambda '} (b) \sum_{\lambda} C_{\lambda} \Psi_{\lambda}( b t^{\Lambda '} ) \psi_{\lambda}(\vec{b}) \psi_{\lambda}(\vec{c}) \ , \quad  \Lambda ' = [2,2] \ ,
\ee 
where $b$ and $w$ are $U(1)_b$ and $SU(2)_w$ fugacities coming from the puncture specified by $\Lambda$.
The former $U(1)_b$ did not appear in the previous expressions because this together with two $SU(N)$'s becomes the $SU(2N)$ group.
A little thought reveals that $ \Psi_{\lambda}( b  t^{\Lambda '} )=\Psi_{\lambda}( (b, w)  t^\Lambda )|_{w\rightarrow t^{\half}}$. From this we see that the indices of $E_7$ and $R_{0, 4}$ are related by
\be 
\label{eq:TowardsE7}
\mathcal{I}_{E_7} = \lim_{w\to t^{\half}} \left(\frac{K_{ \Lambda '} (b)}{ K_\Lambda (b, w)} \mathcal{I}_{R_{0, 4}} (b,w)\right) .
\ee 

This procedure can be thought of as picking up the residue at $w=t^{\half}$ in superconformal index of $R_{0,N}$ (as described in \cite{Gaiotto:2012xa}) and removing the Goldstone modes after Higgsing. Here, upon Higgsing, the $SU(2)$ part of the conserved current is being removed. To this end, we simply get
\begin{align}
 \CI_{E_7} = \lim_{w \to t^{1/2}} \left( \frac{K_{\varnothing}}{K_{SU(2)}(w)} \CI_{R_{0, 4}} (w) \right) \ , 
\end{align}
with
\begin{align}
 K_{SU(2)}(w) = \PE \left[ \frac{t - pq}{(1-p)(1-q)} (w^2 + 1 + w^{-2}) \right] \ , ~~ K_{\varnothing} = \PE \left[ \frac{t^2 - pq t }{(1-p)(1-q)} \right] \ . 
\end{align}
Upon explicitly computing this expression, we find that
\begin{align} 
\begin{split}
\CI_{E_7} &= \frac{\k}{2}  \Gamma(t^2) \Gamma(pqt) \Gamma \left(\frac{pq}{t}\right) \oint_{C_s} \frac{ds}{2\pi i s} \frac{\G(s^{\pm 1}) \G(t^{-1} s^{\pm 1}) }{ \G(s^{\pm 2})} \CI_{SU(4)}(s) \\
 &=\frac{\k}{2} \frac{\G(\frac{\ft^8}{v^2}) \G(\ft^2 v)}{\G( \frac{v}{\ft^4})}  \oint_{C_s} \frac{ds}{2\pi i s} \frac{\G(s^{\pm 1}) \G(\frac{v}{\ft^4} s^{\pm 1}) }{ \G(s^{\pm 2})} \CI_{SU(4)}(s)   \ , 
\end{split}
\end{align}
which can be expanded in $\ft$ to give
\be
\label{eq:E7index}
\begin{split}
	\mathcal{I}_{E_7} &=1+\chi^{E_7}_{133} \frac{ \ft^4}{v}-(\chi^{E_7}_{133}+1) \ft^6+(\chi^{E_7}_{133}+1) \chi^{SU(2)}_2(y)\frac{ \ft^7 }{v} \\
	&~~+\ft^8 \left(\frac{\chi_{7371}^{E_7}}{v^2}+v+v^4\right)+\ft^9 \left(-(\chi_{133}^{E_7} +2 )\chi^{SU(2)}_2(y)-v^3 \chi ^{SU(2)}_2(y)\right)\\
	&~~+\ft^{10} \left(-\frac{\chi^{E_7}_{8645}+\chi^{E_7}_{7371}+\chi^{E_7}_{133}}{v}+v^2+\frac{(\chi^{E_7}_{133}+1)
		\chi ^{SU(2)}_3(y)+1}{v}\right)\\
	&~~+\ft^{11} \left(\frac{\chi^{E_7}_{8645}+\chi^{E_7}_{7371}+\chi^{E_7}_{133} }{v^2}+v +v^4 \right)\chi ^{SU(2)}_2(y)\\
	&~~+\ft^{12} \left(\chi^{E_7}_{7371}+\chi^{E_7}_{1539}+\frac{\chi^{E_7}_{238602}}{v^3}- v^3-(\chi^{E_7}_{133}+2 + v^3) \chi ^{SU(2)}_3(y)\right) + \hdots \ ,
\end{split}
\ee
where, $\chi_d^{E_7}$ denotes the characters of $d$-dimensional irreducible representation of $E_7$ and $\chi^{SU(2)}_n(y)$ denotes the $n$-dimensional irreducible representation of $SU(2)_1 \subset SO(3,1)_{\rm{Lorentz}}$ that commutes with the supercharges used to define the index in \eqref{eq:indexDef}. The appearance of $\chi_{133}^{E_7}$ as the coefficient of $t=\frac{\ft^4}{v}$ shows that there is a conserved current of the $E_7$ flavor symmetry. The coefficient of $\ft^6$ captures the contribution from the conserved flavor currents as well as the stress-energy tensor. 

The coefficients of $t^2=\frac{\ft^8}{v^2}$ and $t^3=\frac{ \ft^{12}}{v^3}$ are given by the characters of the $E_7$ representations with Dynkin labels $[2,0,0,0,0,0,0]$ and $[3,0,0,0,0,0,0]$ respectively.  It follows that our computation reproduces the first three non-trivial terms in the Hall-Littlewood limit of the index ($p \to 0, q \to 0$) for the $E_7$ theory. The Hall-Littlewood index can be identified with the Hilbert series of the Higgs branch. Since our theory realizes the 1-instanton moduli space of $E_7$, the Hall-Littlewood index should be given as \cite{Benvenuti:2010pq, Keller:2011ek,Keller:2012da}
\begin{align}
 \CI_{HL} = \sum_{n \ge 0} \chi_{[n, 0, 0, 0, 0, 0, 0]}^{E_7} t^n \ , 
\end{align}
which is consistent with our result. The various terms appearing in the $E_7$ superconformal index up to $\mathcal{O}(t^{20})$ can be found in appendix \ref{app:IE7}. 

One can also take the Schur limit ($p \to 0, t \to q$) of \eqref{eq:E7index} to get 
\be
\label{eq:E7Schurindex}
\begin{split}
\mathcal{I}_{E_7}^{\rm Schur}&=1+\chi^{E_7}_{133} q+(\chi^{E_7}_{7371}+\chi^{E_7}_{133}+1)q^2\\
&~~+(\chi^{E_7}_{238602}+\chi^{E_7}_{8645}+\chi^{E_7}_{7371}+\chi^{E_7}_{133})q^3 + \mathcal{O}(q^4)
\end{split}
\ee 
This result agrees with the computation done using the TQFT description \cite{Gadde:2011ik,Gadde:2011uv, Agarwal:2013uga}.

\section{Elliptic genus of the $E_7$ instanton string}
\label{sec:2d}

\paragraph{$\CN=(0, 4)$ duality}
It was shown in \cite{Putrov:2015jpa}, that there exists a class of 2d $\CN=(0,4)$ theories that admit duality transformations akin to the dualities of 4d $\mathcal{N}=2$ class-$\CS$ theories. More precisely, the aforementioned 2d $\CN=(0,4)$ theories arise from dimensionally reducing the 4d class-$\CS$ theories on a $\mathbb{CP}^1$ with a partial topological twist. For the Lagrangian theories, the reduction simply maps 4d hypermultiplets to 2d $(0, 4)$ hypermultiplets, and 4d vector multiplets to 2d $(0, 4)$ vectors. It therefore follows that the 2d $\CN=(0,4)$ Lagrangian consisting of an $SU(N)$ gauge theory coupled to $2N$ fundamental $(0,4)$ hypers admits an S-duality frame where it is described by the dimensionally reduced version of $R_{0,N}$ coupled to a doublet of hypermultiplets via an $SU(2)$ gauge group. 

When a 2d $\CN=(0, 2)$ superconformal theory admits a gauge theory description, one can compute the elliptic genus using localization \cite{Benini:2013xpa,Gadde:2013ftv}.
It was shown in \cite{Putrov:2015jpa} that the elliptic integrals that appear in the elliptic genus computation admit an inversion formula which is similar in spirit to the Spiridonov-Warnaar inversion formula for the elliptic beta integrals. This can then be used to extract the elliptic genus of $R_{0,N}$ starting from that of the $SU(N)$ gauge theory with $2N$ fundamental hypers. 

Let us briefly summarize this procedure. The statement of the elliptic inversion formula is: Let $f(z)$ is a meromorphic function such that $f(z)=f(1/z)$ and $f(q z) = q^2 z^4 f(z) $.
Then the following identity holds
\begin{equation}
\frac{(q;q)^4}{4}\int\limits_\mathrm{JK}\frac{d\xi}{2\pi i\xi}\,\int\limits_\mathrm{JK}\frac{d\zeta}{2\pi i\zeta}\,\theta(\xi^{\pm 2})\,\theta(\zeta^{\pm 2})
\,\frac{\theta(v^{-2})}{\theta(v^{-1}\,z^{\pm 1}\xi^{\pm 1})}
\,\frac{\theta(v^{2})}{\theta(v\,\xi^{\pm 1}\zeta^{\pm 1})}f(\zeta)=
f(z) \ .
\label{inv-formula}
\end{equation}
Here $\theta(x)$ is defined as
\be
\theta(x) := (x;q)(q/x;q) \ , \ \ \ (x;q) := \prod_{i=0}^{\infty}(1-x q^i) \ .
\ee  
S-duality implies that elliptic genus of the $SU(N)$ gauge theory is related to the elliptic genus of $R_{0,N}$ by the following 2d version of \eqref{eq:Sduality}:
\be
\label{eq:2dSduality}
\mathcal{I}^{T^2 \times S^2}_{SU(N)} (s, \vec{a}) = \frac{(q;q)^2}{2}\int\limits_\mathrm{JK} \frac{d w}{2 \pi i w} \frac{ \theta(v^2) \theta \left(v^2 w^{\pm 2}\right) \theta(w^{\pm 2}) }{\theta(v\hspace{1pt} w^{\pm 1} s^{\pm 1})} \mathcal{I}^{T^2 \times S^2}_{R_{0, N}} (w , \vec{a} ) \ .
\ee  

If we now choose $f(z)$ to be given by 
\be
f(z) = \frac{\theta(v^{2} z^{\pm 2})}{\theta(v^{-2})}\mathcal{I}^{T^2 \times S^2}_{R_{0, N}} (z , \vec{a} )  \ ,
\ee 
and subsitute in \eqref{eq:2dSduality}, we obtain the following integral expression of the elliptic genus of $R_{0,N}$:
\begin{equation}
\mathcal{I}^{T^2 \times S^2}_{R_{0, N}} (\zeta , \vec{a} ) =\frac{(q;q)^2}{2\,\theta(v^2\zeta^{\pm 2})}
\int\limits_\text{JK}\frac{d s}{2\pi i\,s}\,
\frac{\theta(s^{\pm 2})\theta(v^{-2})}{\theta(v^{-1}s^{\pm 1}\zeta^{\pm 1})}\,
\mathcal{I}^{T^2 \times S^2}_{SU(N)} (s , \vec{a} )  \,.
\label{T3-index-formula}
\end{equation}
Here $\vec{a}$ are the fugacities for the $SU(2N)$ flavor symmetry carried by the $SU(N)$ gauge theory as well as the $R_{0,N}$ theory. When $N=3$, this gives the elliptic genus of Minahan-Nemeschansky's $E_6$ theory \cite{Putrov:2015jpa}.  

We can now use the above identity to obtain the elliptic genus of $R_{0,4}$ theory and then appropriately ``Higgs'' it to obtain the elliptic genus of Minahan-Nemeschansky's $E_7$ theory.

\paragraph{Elliptic genus of the $E_7$ instanton string}

Repeating the same procedure as in the case of 4d, it is straight-forward to write down the 2d $\CN=(0, 2)$ theory from the dimensional reduction of the 4d matter content given in table \ref{tab:E7Lag}. Basically the dimensional reduction maps 4d chiral multiplets to 2d $(0, 2)$ chiral multiplets, but when a 4d chiral multiplet has $U(1)_r$ charge 2, it becomes a 2d $(0, 2)$ Fermi multiplet. 

The elliptic genus can be obtained via following integral:
\begin{align}
 \CI_{E_7}^{T^2 \times S^2} = \frac{(q;q)^2 \theta(v^2)\theta(v^{-2})}{2 \theta(v^4)}\int_{\textrm{JK}} \frac{ds}{2\pi i s} \frac{\theta(s^{\pm 2})}{\theta(s^{\pm 1}) \theta(v^{-2} s^{\pm 1})} \CI_{SU(4)}^{T^2\times S^2} (s) .
\end{align}
The $\theta$ functions in the denominators come from chiral multiplets and the numerators come from Fermi multiplets. One see that this expression can be obtained via simply taking $\G(t^T z) \to 1/\theta(v^{2T} z)$ from the integral expression for the 4d index \eqref{eq:E7index}. 
The integration contour is chosen by the Jefferey-Kirwan formula \cite{Benini:2013xpa}. Evaluating the integral, we obtain a large expression in terms of ratio of theta functions. Upon expanding in terms of $q$ and $v$, we obtain
\begin{align}
\begin{split}
 \CI_{E_7}^{T^2 \times S^2} &= 1 + \chi^{E_7}_{133} v^2 + \chi^{E_7}_{7371}v^4 + \chi^{E_7}_{238602} v^6 + \ldots \\
 &\quad+ q \big( 1 + \chi^{E_7}_{133} + \left( 1 + 2 \chi^{E_7}_{133} + \chi^{E_7}_{7331} + \chi^{E_7}_{8645}  \right) v^2 \\
  &\quad\quad\quad + \left( \chi^{E_7}_{133} + 2  \chi^{E_7}_{7371} + \chi^{E_7}_{8645} + \chi^{E_7}_{238602} + \chi^{E_7}_{573440} \right) v^4+ \ldots \big) \\
  &\quad+ q^2 \left( 3 + 2  \chi^{E_7}_{133} + \chi^{E_7}_{1539} + \chi^{E_7}_{7371} + \ldots \right) \\
  &\quad+ \CO(q^3) \ . 
\end{split}
\end{align}
The $q^0$ term reproduces the Hilbert series of the (centered) 1-instanton moduli space of $E_7$ \cite{Benvenuti:2010pq,Keller:2011ek, Keller:2012da}, as it should. The stringy contributions agree with the result of \cite{DelZotto:2016pvm}. 

\section{Conclusion and Discussion}
In this paper, we have obtained an $\CN=1$ Lagrangian theory that flows to the Minahan-Nemeschansky $\CN=2$ SCFT with $E_7$ global symmetry following a method that is very similar the one proposed for the $E_6$ theory \cite{Gadde:2015xta}. Along with the ``Lagrangian" for the $E_6$ theory, our version shares an identical issue of not having manifest $SU(2)_s$ gauge symmetry that we gauge. This symmetry is visible in the S-dual frame, but not in the ``electric" frame. Even though our description has this short coming, it was enough for us to utilize it to compute various supersymmetric partition functions, some of which have not been obtained previously. 

The method presented in the current paper (which is originated from \cite{Gadde:2015xta}) can be used whenever some dual description of an SCFT has $SU(2)$ gauge group coupled with 2 fundamental chiral multiplets (=1 full hyper). Or from the class $\CS$ perspective, any theory that has the puncture of type $[N-2, 1^2]$ or $[N-2, 2]$. Unfortunately, we are not aware of any such dual frames involving the $E_8$ SCFT. From the class $\CS$ viewpoint, this theory is realized via 3-punctured sphere of $A_5$ theory with puncture types $[1^6], [2^3], [3^2]$. One can easily obtain 2 out of 3 punctures from the S-dual description of the $SU(6)$ SQCD upon Higgsing one of the $[1^6]$ punctures, but not all 3 of them. It would be very interesting to find a way to obtain the puncture of type $[N-3, 1^3]$ or $[N-4, 2^2]$, which would be enough for us to get the $E_8$ theory. 

One of the applications of our ``Lagrangian" description is on the instanton strings that appear in the ``atomic classification" of 6d $\CN=(1, 0)$ theories \cite{Heckman:2013pva, Heckman:2015bfa}. Recently, ADHM-like construction for $G_2$ theories with fundamental matter multiplets have been found \cite{Kim:2018gjo} (see also \cite{Kim:2016foj} for the case of $SU(3)$). Our construction significantly differs from their gauge theory attempts for the $E_7$, and it would be interesting to see if our construction can be related to (a modification of) theirs so that it can be generalized to arbitrary number of instantons. Even though such a construction exists in class $\CS$ \cite{Benini:2009gi, Gaiotto:2012uq}, we are not aware of a gauge theory realization which would enable us to compute the elliptic genus of the instanton strings of $E_7$ or $E_8$.

\acknowledgments
While the current paper was in its final stage, we have been informed that a similar result for the superconformal index has been also obtained by Hee-Cheol Kim, Shlomo Razamat and Gabi Zafrir. We thank them for sharing this information. 
We would like to thank Amihay Hanany, 
Seok Kim and Kimyeong Lee 
for useful discussions. 
KM and JS would like to thank KEK Theory Group for hospitality during the East Asia Joint Workshop on Fields and Strings 2017, where this project was conceived. 
The work of PA is supported in part by Samsung Science and Technology Foundation under Project Number SSTF-BA1402-08, in part by National Research Foundation of Korea grant number 2015R1A2A2A01003124 and in part by the Korea Research Fellowship Program through the National Research Foundation of Korea funded by the Ministry of Science, ICT and Future Planning, grant number 2016H1D3A1938054.
The work of KM is supported by JSPS KAKENHI Grant Number JP17K14296.
The work of JS is supported in part by the Overseas Research Program for Young Scientists through Korea Institute for Advanced Study (KIAS).

\appendix

\section{More on $R_{0,N}$ theory}
  We consider the chiral ring of the $R_{0,N}$ theory from the two different perspectives.
  In section \ref{subsec:duality}, we see the chiral ring described in section \ref{subsec:R0NLag} is consistent with the duality of the SQCD.
  In section \ref{subsec:TN}, some chiral ring relations are shown to be obtained from those of the $T_N$ theory by nilpotent Higgsing.

\subsection{On the duality of SQCD}
\label{subsec:duality}
  We discuss the duality of $\CN=2$ $SU(N)$ SQCD with $2N$ flavors.
  As found in \cite{Gaiotto:2009we, Chacaltana:2010ks}, this theory is dual to $R_{0,N}$ where the $SU(2)$ flavor group
  is gauged with a fundamental hypermultiplet, $\mathfrak{q}$ and $\tilde{\mathfrak{q}}$. 
  The global symmetry is $SU(2N) \times U(1)_s$ in addition to $U(2)_R$ symmetry.
  
\paragraph{Electric side}
  Let us first see the chiral ring of the electric side.
  We denote the $2N$ hypermultiplets as $Q^f$ and $\tilde{Q}_f$. 
  The ring is simply generated by mesons $M^f_{f'}$, baryons $B^{f_1,f_2,\ldots,f_N}$, anti-baryons $\bar{B}_{f_1,f_2,\ldots,f_N}$, and
  Coulomb branch operators $\tr \Phi^i$,
  where $M^f_{f'}=Q^f \tilde{Q}_{f'}$ and $B^{f_1,f_2,\ldots,f_N} = Q^{f_1} Q^{f_2} \cdots Q^{f_N}$ 
  where the gauge indices are contracted. 
  We divide the meson into the traceless part and the singlet:
    \bea
    \hat{M}_{f'}^f
     =     M_{f'}^f - \frac{1}{2N} (\Tr M) \delta^f_{f'}, ~~~
    M_0
     =     \Tr M
    \eea
  See table \ref{tab:electric} for the flavor charges of them.
  We normalize the baryonic $U(1)_s$ charges of baryons to be $\pm 1$.
    \begin{table}
    \begin{center}
	\begin{tabular}{|c||c|c|c|c|}
		\hline
		 & $SU(2N)$ & $U(1)_s$ & $U(1)_{I_3}$ & $U(1)_{r}$ \\ 
		\hline \hline
		$\hat{M}$ & $\mathbf{adj}$ &$0$ & $1$ & 0 \\
	        $M_0$ & $\mathbf{1}$ &$0$ & $1$ & 0 \\
		$B$ & $\mathbf{\wedge}^N$ & $1$ & $\frac{N}{2}$ & 0 \\
		$\bar{B}$& $\bar{\mathbf{\wedge}}^N$ & $-1$ & $\frac{N}{2}$ & 0\\ 
		$\tr \Phi^i$ & $\mathbf{1}$ & $0$ & $0$ & $2i$ \\
		\hline
		\end{tabular}
		\caption{Charges of the gauge invariant operators of $\CN=2$ SQCD with $2N$ flavors.} 
		\label{tab:electric}   
    \end{center}
    \end{table}
  The chiral ring relations are
    \bea
    M^{[ f_1}_{f'_1} M^{f_2}_{f'_2} \cdots M^{f_N]}_{f'_N}
    &=&    \frac{1}{2N} B^{f_1, f_2,\ldots f_N} \bar{B}_{f'_1,\ldots, f'_N},
               \label{relation1SQCD} \\
    \hat{M}^f_{f'} \hat{M}^{f'}_{f''}
    &=&    \frac{1}{4N^2} M_0^2 \delta^f_{f''},
               \label{chiral2SQCD} \\
    \hat{M}^f_{f'} B^{f',f_2,\ldots,f_N}
    &=&    \frac{1}{2N} M_0 B^{f,f_2,\ldots,f_N},
               \label{relation3SQCD} \\
    \hat{M}^f_{f'} \bar{B}_{f,f_2,\ldots,f_N}
    &=&    \frac{1}{2N} M_0 \bar{B}_{f',f_2,\ldots,f_N},
               \label{relation4SQCD} \\
    X \tr \Phi^i
    &=&    0,
    \eea
  where $X$ is arbitrary Higgs branch operator.

\paragraph{Magnetic side}
  On the magnetic side, the $SU(2)$ gauge theory has a superpotential
    \bea
    W 
     =     \tr \phi_D \mu_{SU(2)} + \tr \phi_D \mathfrak{q} \tilde{\mathfrak{q}},
    \eea
  where the gauge $SU(2)$ indices are contracted properly. 
  The F-term relations are 
    \bea
    0
     =     (\mu_{SU(2)})^\alpha_{~\beta} + \mathfrak{q}^\alpha \tilde{\mathfrak{q}}_\beta - \half \mathfrak{q}^\gamma \tilde{\mathfrak{q}}_{\gamma} \delta^\alpha_\beta, ~~~~
    0
     =     \phi_D \mathfrak{q}, ~~~~
    0
     =     \tilde{\mathfrak{q}} \phi_D.
    \eea
  From these, any gauge invariant operator involving $\mu_{SU(2)}$ can be written in terms of $\mathfrak{q}$ and $\tilde{\mathfrak{q}}$.
  In particular, $\tr \mu_{SU(2)}^2 = (\Tr \mathfrak{q} \tilde{\mathfrak{q}} )^2$.
  Thus we list the operators constructed from $\mu_{SU(2N)}$, $Q_{(N)}$, $\phi_D$, $\mathfrak{q}$, and $\tilde{\mathfrak{q}}$
  in table \ref{tab:magnetic}.
  Note that as discussed in section \ref{subsec:R0NLag}, the $U(1)_s$ charges of $\mathfrak{q}$ and $\tilde{\mathfrak{q}}$ are normalized to be $\pm 1$.
    \begin{table}
    \begin{center}   
	\begin{tabular}{|c||c|c|c|c|}
		\hline
		 & $SU(2N)$ & $U(1)_s$ & $U(1)_{I_3}$ & $U(1)_{r}$ \\ 
		\hline \hline
		$\mu_{SU(2N)}$ & $\rm{adj}$ &$0$ & $1$ & 0 \\
	        $\mathfrak{q} \tilde{\mathfrak{q}}$ & $\mathbf{1}$ &$0$ & $1$ & 0 \\
	        $Q_{(N)} \mathfrak{q}$ & $\mathbf{\wedge}^N$ & $1$ & $\frac{N}{2}$ & 0 \\
		$\epsilon \cdot Q_{(N)} \tilde{\mathfrak{q}}$& $\bar{\mathbf{\wedge}}^N$ & $-1$ & $\frac{N}{2}$ & 0\\
		$\tr \phi_D^2$ & $\mathbf{1}$ & $0$ & $0$ & $4$ \\
		$u_d$ & $\mathbf{1}$ & $0$ & $0$ & $2d$ \\
		\hline
		\end{tabular}
		\caption{Charges of the gauge invariant operators of the dual theory of SQCD.}  
		\label{tab:magnetic}
    \end{center}
    \end{table}
    
  It is easy to see the agreement with the chiral ring of the electric side:
    \bea
    \hat{M}.    
    &=&    \mu_{SU(2N)},~~~
    M_0
     =     \mathfrak{q}^\alpha \tilde{\mathfrak{q}}_\alpha,
            \nonumber \\
    B^{f_1,\ldots,f_N}
    &=&    Q_{(N)~~~~\alpha}^{f_1,\ldots, f_N} \mathfrak{q}^\alpha, ~~~
    \bar{B}_{f_1,\ldots,f_N}
     =     \epsilon_{f_1,\ldots,f_{2N}} Q_{(N)~~~~~~~~\alpha}^{f_{N+1},\ldots, f_{2N}} \tilde{\mathfrak{q}}_\beta \epsilon^{\alpha \beta}, 
            \nonumber \\
    \tr \Phi^2
    &=&    \tr \phi_D^2, ~~~
    \tr \Phi^k
     =     u_k. ~~~(k \geq 3)
    \eea
  
  Moreover one can identify the chiral ring relation.
  The \eqref{relation1R}  plus the F-term condition leads to 
    \bea
    \mu_{SU(2N)}^2
     =     (\tr q \tilde{q})^2 \mathbf{1}_{2N \times 2N}.
    \eea
  This agrees with \eqref{chiral2SQCD}.
  Then the relation \eqref{relation2R} multiplied by $q^\alpha$ with the F-term condition gives
    \bea
    (\mu_{SU(2N)})^{f_1}_{~f'} Q_{(N)~~~~~~~\alpha}^{f',f_2,\ldots, f_N} q^\alpha
     =     \frac{1}{2N} (q^\gamma \tilde{q}_\gamma) Q^{f_1,f_2,\ldots,f_N}_{(N)~~~~~~~\alpha} q^\alpha,
    \eea
  where we have used the F-term condition.
  This is indeed \eqref{relation3SQCD}.
  The similar analysis for \eqref{relation3} leads to \eqref{relation4SQCD}.

\subsection{From $T_N$ chiral ring}
\label{subsec:TN}
  One could study the chiral ring of the $R_{0,N}$ from that of the $T_N$ theory.
  The latter has been found in \cite{Maruyoshi:2013hja} to be
    \begin{itemize}
    \item Moment map operators, $\mu_A$, $\mu_B$, and $\mu_C$, 
             transforming in the adjoint representations of $SU(N)_A$, $SU(N)_B$ and $SU(N)_C$ respectively.
             The $(I_3, r)$ charge of them are $(1, 0)$, thus the dimensions are $\Delta=2$.
    \item $Q_{(k)}$, where $k=1,\ldots,N-1$, 
            transforming in $(\wedge^k, \wedge^k, \wedge^k)$ representations of $SU(N)_A \times SU(N)_B \times SU(N)_C$.
            These have charge $(\frac{k(N-k)}{2}, 0)$.
    \item The Coulomb branch operators, $u_{d, i}$, where $d=3,4,\ldots,N$ and $i=1,\ldots, d-2$,
            with $(I_3, r) = (0, 2d)$ and $\Delta= d$.
    \end{itemize}
  See \cite{Lemos:2014lua} for further investigation.
  There are chiral ring relations among them, for example \cite{Maruyoshi:2013hja},
    \bea
    \tr \mu_A^k
    &=&    \tr \mu_B^k
     =     \tr \mu_C^k,
             \label{relation1} \\
    (\mu_A)^{i_A}_{j_A} Q_{(1)}^{j_A, i_B, i_C}
    &=&    (\mu_B)^{i_B}_{j_B} Q_{(1)}^{i_A, j_B, i_C}
     =     (\mu_C)^{i_C}_{j_C} Q_{(1)}^{i_A, i_B, j_C},
            \label{relation2} \\
    Q_{(1)}^{i_A, I_B, i_C} Q_{(N-1) j_A, j_B, i_C}
    &=&    \sum_{l=0}^N  v_l  \sum_{m=0}^{N-l-1} (\mu_A^{N-l-1-m})^{i_A}_{~j_A} (\mu_B^m)^{i_B}_{~j_B}, 
               \label{relation3} 
    \eea
    \bea
    &\frac{1}{(N-1)!}Q_{(1)}^{i_{A,1} i_{B,1} i_{C,1}}Q_{(1)}^{i_{A,2} i_{B,2} i_{C,2}}\cdots Q_{(1)}^{i_{A,N-1} i_{B,N-1} i_{C,N-1}}
    \epsilon_{ i_{B,1}i_{B,2}\cdots i_{B,N-1}i_B }\epsilon_{ i_{C,1}i_{C,2}\cdots i_{C,N-1} i_C} \nonumber \\
    &= Q_{(N-1)i_A i_B i_C} (\mu_A^0)^{(i_{A,1}}_{~j_{A,1}} (\mu_A)^{i_{A,2}}_{~j_{A,2}} (\mu^2_A)^{i_{A,3}}_{~j_{A,3}}\cdots
    (\mu^{N-2}_A)^{i_{A,N-1})}_{~j_{A,N-1}}
    \epsilon^{  j_{A,1} j_{A,2} j_{A,3}\cdots   j_{A,N-1} i_A}, 
    \label{relation4}
    \eea
    \bea
    &\frac{1}{(N-1)!}Q_{i_{A,1} i_{B,1} i_{C,1}}Q_{i_{A,2} i_{B,2} i_{C,2}}\cdots Q_{i_{A,N-1} i_{B,N-1} i_{C,N-1}}
    \epsilon^{ i_{B,1}i_{B,2}\cdots i_{B,N-1} i_B}\epsilon^{ i_{C,1}i_{C,2}\cdots i_{C,N-1} i_C} \nonumber \\
    &=(-1)^{\frac{1}{2}N(N-1)} Q^{i_A i_B i_C} (\mu^0_A)^{j_{A,1}}_{~(i_{A,1}} (\mu_A)^{j_{A,2}}_{~i_{A,2}}(\mu^2_A)^{j_{A,3}}_{~i_{A,3}}
    \cdots (\mu^{N-2}_A)^{j_{A,N-1}}_{~i_{A,N-1})}
    \epsilon_{  j_{A,1} j_{A,2}\cdots   j_{A,N-2} j_{A,N-1} i_A},
    \label{relation5}
    \eea
  where $v_l$ are defined by
    \bea
    P_X(x)
     =     \det(x {\bf 1}-\mu_X)
     =     \sum_{k=0}^N v_{X,k} x^{N-k}.
    \eea
  There are also relations obtained by applying  the cyclic permutation $A \rightarrow B \rightarrow C \rightarrow A$ to the above relations.
  There should be other equations involving $Q_{(k)}$ with $k=2,\ldots,N-2$, however we do not know the form of the relations.
  
  We start from these and see what happens if we close the $SU(N)_C$ puncture.
  If we represent the Higgsing by the partition $N = \sum_k k n_k$, the corresponding puncture is represented by $n_1 = 2$, and $n_{N-2} =1$.
  Let us first consider the operator $\mu_C$ which are in the adjoint representation of $SU(N)_C$.
  This representation is decomposed into
    \bea
    {\rm adj}
     \rightarrow
           V_{\frac{N-3}{2}} \otimes (\mathbf{2} \oplus \bar{\mathbf{2}}) \oplus V_0 \otimes (\mathbf{3})
           \oplus \oplus_{k=1}^{N-2} V_{k-1} \otimes \mathbf{1},
    \eea
  where $V_s$ is the spin-$s$ representation of $SU(2)$.
  As the components of $\mu_C$, only $\sigma_3 = -j$ component of each representation survives. 
  Therefore, the moment map operator $\mu_C$ can be written
    \bea
    \mu_C
    &\rightarrow&
           \mu_{\frac{N-3}{2}, - \frac{N-3}{2}, i} T_i + \tilde{\mu}_{\frac{N-3}{2}, - \frac{N-3}{2}, i} \tilde{T}_i + \mu_{SU(2)} T_{SU(2)} + \mu_{U(1)} T_{U(1)} + \sum_{k=2}^{N-2} \mu_{k-1,-k+1} T_k
           \nonumber \\
    &=&     \left( \begin{array}{cccc|cc}
           \mu_{U(1)} & (\rho(\sigma^+))^1_{~2}  & 0 & 0&0 & 0 \\ 
           \vdots & \ddots &  \ddots & 0 & \vdots & \vdots \\ 
           \mu^{N-4,-(N-4)} &    & \ddots & (\rho(\sigma^+))^{N-3}_{~N-2}  & 0 & 0 \\
           \mu^{N-3,-(N-3)} &\mu^{N-4,-(N-4)}  & \ldots& \mu_{U(1)} & \mu^{\frac{N-3}{2},-\frac{N-3}{2}, 1} & \mu^{\frac{N-3}{2},-\frac{N-3}{2}, 2}
           \\ \hline
           \tilde{\mu}^{\frac{N-3}{2},-\frac{N-2}{2},1} & 0 & \ldots & 0 & -\frac{N-2}{2} \mu_{U(1)} + J_3 & J_+  \\
           \tilde{\mu}^{\frac{N-3}{2},-\frac{N-2}{2},2} & 0 & \ldots & 0 & J_- &  -\frac{N-2}{2} \mu_{U(1)} - J_3 \\
           \end{array} \right),
           \label{eq:matrixformA}
    \eea
  where $\mu_{\frac{N-3}{2}, - \frac{N-3}{2}, i}$ and $\tilde{\mu}_{\frac{N-3}{2}, - \frac{N-3}{2}, i}$ are in $\mathbf{2}$ and $\bar{\mathbf{2}}$ 
  representations of $SU(2)$ respectively, and
  $J_3$, $J_{\pm}$ forming $\mu_{SU(2)}$ are SU(2) moment map operators.
  Since the IR $U(1)_{I_3}$ symmetry which is unbroken is
    \bea
    I_{3, IR}
     =     I_3 - \rho(\sigma_3),
    \eea
  one can see that the operators $\mu_{SU(2)}$ and $\mu_{U(1)}$ have charge one under this.
  By using the chiral ring relation \eqref{relation1}, the operators $\mu_{k-1, -k+1}$ with $k=2,\ldots,N-2$ can be written in terms of 
  $\mu_A$, $\mu_B$, $\mu_{SU(2)}$, $\mu_{U(1)}$ $\mu_{\frac{N-3}{2}, - \frac{N-3}{2}, i}$ and $ \tilde{\mu}_{\frac{N-3}{2}, - \frac{N-3}{2}, i}$.
    
  We now then turn to the operator $Q_{(1)}$ or $Q_{(N-1)}$ in $T_N$.
  Since these are in the fundamental or the anti-fundamental representation of $SU(N)_C$, one needs the decomposition of this:
    \bea
    \mathbf{N}
     \rightarrow
          V_{\frac{N-3}{2}} \otimes \mathbf{1} \oplus V_0 \otimes \mathbf{2}.
    \eea
  Thus, we write $Q_{(1)}$ as
    \bea
    Q_{(1)}
     =     Q_{m}^{i_A, i_B} + Q^{i_A, i_B, i}
    \eea
  where $i=1,2$ are $SU(2)$ indices, and $m=0,1,\ldots,N-3$. 
  The charges of $Q_{m}^{i_A, i_B}$ and $Q^{i_A, i_B, i}$ are $(I_3, r) = (N-2-m, 0)$, and $(\frac{N-1}{2},0)$ respectively.
  The similar decomposition can be obtained from $Q_{(N-1)}$. 
  Note that there are $I_3=1$ operator $Q_{N-3}^{i_A, i_B}$ from $Q_{(1)}$ and $\bar{Q}_{N-3, i_A, i_B}$ from $Q_{(N-1)}$.
  These transform in $(\mathbf{N}, \mathbf{N})$ and $(\bar{\mathbf{N}}, \bar{\mathbf{N}})$ representation of $SU(N)_A \times SU(N)_B$.
  In total, we have $I_3=1$ operators:
    \bea
    \mu_A,~~ \mu_B, ~~Q_{N-3}^{i_A, i_B}, ~~\bar{Q}_{N-3, i_A, i_B}, ~~ \mu_{U(1)}
    \eea
  in addition to $\mu_{SU(2)}$. 
  These form the $SU(2N)$ adjoint representation, thus are interpreted as the moment map operator of $SU(2N)$ flavor symmetry of $R_{0,N}$.
  Furthermore, by using \eqref{relation2}, other component of $Q_{m}^{i_A, i_B}$ can be written 
  in terms of $\mu_A$, $\mu_{U(1)}$, $\mu_{SU(2)}$ and $Q_{N-3}^{i_A, i_B}$.
  Thus these are not the generators of the chiral ring.
  
  Moreover, as discussed in \cite{Collinucci:2017bwv}, the relation \eqref{relation3} leads to \eqref{relation1R}
    \bea
    \mu_{SU(2N)}^2
     \sim     \tr \mu_{SU(2)}^2 \mathbf{1}_{2N\times 2N}.
    \eea
  
  Let us then see the other operators $Q_{(k)}$, which is in the $\wedge^k$ representation of $SU(N)_C$.
  We will not fully work out the Higgsing, instead just see that to each decomposition of $\wedge^k$, 
  there appear the representation $\mathbf{2}$ of $SU(2)$ which multiplied by $\otimes V_{s_i}$ representation.
  The highest spin $s_{max} = \frac{(k-1)(N-k-1)}{2}$.
  For example, one can explicitly calculate that
    \bea
    \wedge^2
    &\rightarrow&
         V_{\frac{N-3}{2}} \otimes \mathbf{2} \oplus \ldots, ~~~
         \nonumber \\
    \wedge^3
    &\rightarrow&
         V_{N-4} \otimes \mathbf{2} \oplus \ldots, 
    \eea
  and so on.
  Since the largest spin representation has highest $\rho(\sigma_3)$, this gives the operator with the smallest $I_3$ charge,
  which is $I_3 = \frac{N-1}{2}$ for any $k$.
  Therefore we have operators which are in $(\wedge^k, \wedge^k, \mathbf{2})$ with $k=2,3,\ldots,N-2$, with $I_3 = \frac{N-1}{2}$.
  In addition to these, there are $Q^{i_A, i_B, i}$, $\bar{Q}_{i_A, i_B}^i$ and $\mu_{\frac{N-3}{2},-\frac{N-3}{2},i}$, 
  and $\tilde{\mu}_{\frac{N-3}{2},-\frac{N-3}{2},i}$, 
  which are in $(\wedge, \wedge, \mathbf{2})$, $(\wedge^{N-1}, \wedge^{N-1}, \mathbf{2})$, $(1,1, \mathbf{2}))$ and $(1,1, \mathbf{2}))$
  representations respectively, with the same $I_3 =\frac{N-1}{2}$ charge.
  Since 
    \bea
    \wedge^N_{SU(2N)}
     \rightarrow     \oplus_{k=0}^{N} (\wedge^k_{SU(N)_A}, \wedge^k_{SU(N)_B}),
    \eea
  all those operators become the $(\wedge^N, \mathbf{2})$ representation of $SU(2N) \times SU(2)$.
  We denote this as $Q_{(N)}$.

\section{Higher order terms for the $R_{0, 4}$ superconformal index}\label{app:IR4}
Here we give an explicit expression for the superconformal index of the $R_{0, 4}$ theory to $\CO(t^{19})$. The characters $\chi_n^w$ denotes that of $n$-dimensional representation of $SU(2)_w$ symmetry, whereas $\chi_n^y$ is for the $SU(2)_1$ symmetry of the Lorentz group. We turned off other flavor fugacities. 
\begin{align*}
\mathcal{I}_{R_{0, 4}} (w) &= 1+\ft^4 \left(\frac{63}{v}+\frac{\chi ^w_3}{v}\right)+\ft^6 \left(-64+v^3+\frac{70 \chi ^w_2}{v^{3/2}}-\chi ^w_3\right)+\ft^7 \left(\frac{64 \chi ^y_2}{v}-v^2 \chi ^y_2+\frac{\chi ^w_3 \chi
	^y_2}{v}\right)\\
&+\ft^8 \left(\frac{1953}{v^2}+2 v+v^4-\frac{70 \chi ^w_2}{\sqrt{v}}+\frac{63 \chi ^w_3}{v^2}+\frac{\chi ^w_5}{v^2}\right)+\ft^9 \left(-65 \chi ^y_2+\frac{70 \chi ^w_2
	\chi ^y_2}{v^{3/2}}-\chi ^w_3 \chi ^y_2\right)\\
&+\ft^{10} \left(-\frac{3968}{v}+63 v^2+\frac{3654 \chi ^w_2}{v^{5/2}}-\frac{128 \chi ^w_3}{v}+\frac{70 \chi ^w_4}{v^{5/2}}-\frac{\chi
	^w_5}{v}+\frac{64 \chi ^y_3}{v}-v^2 \chi ^y_3+\frac{\chi ^w_3 \chi ^y_3}{v}\right)\\
&+\ft^{11} \left(\frac{3969 \chi ^y_2}{v^2}-61 v \chi ^y_2+v^4 \chi ^y_2-\frac{70 \chi ^w_2 \chi
	^y_2}{\sqrt{v}}+\frac{128 \chi ^w_3 \chi ^y_2}{v^2}+\frac{\chi ^w_5 \chi ^y_2}{v^2}\right)\\
&+\ft^{12} \Bigg(2141+\frac{39774}{v^3}-65 v^3+v^6-\frac{8134 \chi ^w_2}{v^{3/2}}+66 \chi
^w_3+\frac{3717 \chi ^w_3}{v^3}+v^3 \chi ^w_3-\frac{140 \chi ^w_4}{v^{3/2}}\\
& \hspace{40pt}+\frac{63 \chi ^w_5}{v^3}+\frac{\chi ^w_7}{v^3}-65 \chi ^y_3+\frac{70 \chi ^w_2 \chi
	^y_3}{v^{3/2}}-\chi ^w_3 \chi ^y_3\Bigg)+\\
&+\ft^{13} \Bigg(-\frac{8128 \chi ^y_2}{v}+191 v^2 \chi ^y_2-v^5 \chi ^y_2+\frac{8134 \chi ^w_2 \chi ^y_2}{v^{5/2}}-\frac{258 \chi ^w_3
	\chi ^y_2}{v}+\frac{140 \chi ^w_4 \chi ^y_2}{v^{5/2}}-\frac{2 \chi ^w_5 \chi ^y_2}{v}\\
&\hspace{40pt}+\frac{64 \chi ^y_4}{v}-v^2 \chi ^y_4+\frac{\chi ^w_3 \chi ^y_4}{v}\Bigg)\\
&+\ft^{14}
\Bigg(-\frac{120772}{v^2}+977 v+v^4+v^7+\frac{94752 \chi ^w_2}{v^{7/2}}+\frac{4550 \chi ^w_2}{\sqrt{v}}-\frac{10035 \chi ^w_3}{v^2}-v \chi ^w_3-v^4 \chi ^w_3\\
&\hspace{40pt}+\frac{3654 \chi
	^w_4}{v^{7/2}}+\frac{70 \chi ^w_4}{\sqrt{v}}-\frac{128 \chi ^w_5}{v^2}+\frac{70 \chi ^w_6}{v^{7/2}}-\frac{\chi ^w_7}{v^2}+\frac{6050 \chi ^y_3}{v^2}-125 v \chi ^y_3+v^4 \chi
^y_3\\
&\hspace{40pt}-\frac{70 \chi ^w_2 \chi ^y_3}{\sqrt{v}}+\frac{192 \chi ^w_3 \chi ^y_3}{v^2}-v \chi ^w_3 \chi ^y_3+\frac{2 \chi ^w_5 \chi ^y_3}{v^2}\Bigg)\\
&+\ft^{15} \Bigg(3182 \chi
^y_2+\frac{122851 \chi ^y_2}{v^3}-66 v^3 \chi ^y_2-v^6 \chi ^y_2-\frac{17234 \chi ^w_2 \chi ^y_2}{v^{3/2}}+70 v^{3/2} \chi ^w_2 \chi ^y_2+133 \chi ^w_3 \chi ^y_2\\
&\hspace{40pt}+\frac{10101
	\chi ^w_3 \chi ^y_2}{v^3}+2 v^3 \chi ^w_3 \chi ^y_2-\frac{280 \chi ^w_4 \chi ^y_2}{v^{3/2}}+\chi ^w_5 \chi ^y_2+\frac{128 \chi ^w_5 \chi ^y_2}{v^3}+\frac{\chi ^w_7 \chi
	^y_2}{v^3}-65 \chi ^y_4\\
&\hspace{40pt}+\frac{70 \chi ^w_2 \chi ^y_4}{v^{3/2}}-\chi ^w_3 \chi ^y_4\Bigg)\\
&+\ft^{16} \Bigg(\frac{599570}{v^4}+\frac{123638}{v}-3116 v^2+65 v^5+v^8-\frac{319718 \chi
	^w_2}{v^{5/2}}-140 \sqrt{v} \chi ^w_2+\frac{119154 \chi ^w_3}{v^4}\\
&\hspace{40pt}+\frac{8395 \chi ^w_3}{v}-v^2 \chi ^w_3-\frac{11718 \chi ^w_4}{v^{5/2}}+\frac{3717 \chi ^w_5}{v^4}+\frac{65 \chi
	^w_5}{v}+v^2 \chi ^w_5-\frac{140 \chi ^w_6}{v^{5/2}}+\frac{63 \chi ^w_7}{v^4}+\frac{\chi ^w_9}{v^4}\\
&\hspace{40pt}-\frac{12289 \chi ^y_3}{v}+256 v^2 \chi ^y_3-v^5 \chi ^y_3+\frac{12684 \chi
	^w_2 \chi ^y_3}{v^{5/2}}-70 \sqrt{v} \chi ^w_2 \chi ^y_3-\frac{388 \chi ^w_3 \chi ^y_3}{v}+v^2 \chi ^w_3 \chi ^y_3\\
&\hspace{40pt}+\frac{210 \chi ^w_4 \chi ^y_3}{v^{5/2}}-\frac{3 \chi
	^w_5 \chi ^y_3}{v}+\frac{64 \chi ^y_5}{v}-v^2 \chi ^y_5+\frac{\chi ^w_3 \chi ^y_5}{v}\Bigg)\\
&+\ft^{17} \Bigg(-\frac{374744 \chi ^y_2}{v^2}+7934 v \chi ^y_2-127 v^4 \chi ^y_2+v^7
\chi ^y_2+\frac{324268 \chi ^w_2 \chi ^y_2}{v^{7/2}}+\frac{9310 \chi ^w_2 \chi ^y_2}{\sqrt{v}}\\
&\hspace{40pt}-\frac{27095 \chi ^w_3 \chi ^y_2}{v^2}+59 v \chi ^w_3 \chi ^y_2-2 v^4 \chi ^w_3
\chi ^y_2+\frac{11788 \chi ^w_4 \chi ^y_2}{v^{7/2}}+\frac{140 \chi ^w_4 \chi ^y_2}{\sqrt{v}}\\
&\hspace{40pt}-\frac{321 \chi ^w_5 \chi ^y_2}{v^2}-v \chi ^w_5 \chi ^y_2+\frac{140 \chi ^w_6
	\chi ^y_2}{v^{7/2}}-\frac{2 \chi ^w_7 \chi ^y_2}{v^2}+\frac{8066 \chi ^y_4}{v^2}-189 v \chi ^y_4+2 v^4 \chi ^y_4\\
&\hspace{40pt}-\frac{70 \chi ^w_2 \chi ^y_4}{\sqrt{v}}+\frac{257 \chi ^w_3
	\chi ^y_4}{v^2}-2 v \chi ^w_3 \chi ^y_4+\frac{2 \chi ^w_5 \chi ^y_4}{v^2}\Bigg)\\
&+\ft^{18} \Bigg(-38573-\frac{2379151}{v^3}+2989 v^3-66 v^6+v^9+\frac{1631264 \chi
	^w_2}{v^{9/2}}+\frac{365050 \chi ^w_2}{v^{3/2}}+140 v^{3/2} \chi ^w_2\\
&\hspace{40pt}-2203 \chi ^w_3-\frac{444555 \chi ^w_3}{v^3}+\frac{119448 \chi ^w_4}{v^{9/2}}+\frac{12544 \chi
	^w_4}{v^{3/2}}+\chi ^w_5-\frac{11799 \chi ^w_5}{v^3}-v^3 \chi ^w_5 +\\
&\hspace{40pt}+\frac{3654 \chi ^w_6}{v^{9/2}}+\frac{70 \chi ^w_6}{v^{3/2}}-\frac{128 \chi ^w_7}{v^3}+\frac{70 \chi
	^w_8}{v^{9/2}}-\frac{\chi ^w_9}{v^3}+1485 \chi ^y_3+\frac{252274 \chi ^y_3}{v^3}-4 v^3 \chi ^y_3\\
&\hspace{40pt}-2 v^6 \chi ^y_3-\frac{26404 \chi ^w_2 \chi ^y_3}{v^{3/2}}+140 v^{3/2} \chi
^w_2 \chi ^y_3+138 \chi ^w_3 \chi ^y_3+\frac{18699 \chi ^w_3 \chi ^y_3}{v^3}+3 v^3 \chi ^w_3 \chi ^y_3\\
&\hspace{40pt}-\frac{420 \chi ^w_4 \chi ^y_3}{v^{3/2}}+\chi ^w_5 \chi
^y_3+\frac{256 \chi ^w_5 \chi ^y_3}{v^3}+\frac{2 \chi ^w_7 \chi ^y_3}{v^3}-65 \chi ^y_5+\frac{70 \chi ^w_2 \chi ^y_5}{v^{3/2}}-\chi ^w_3 \chi ^y_5\Bigg) + \mathcal{O}(\ft^{19})
\end{align*}

\section{Higher order terms for the $E_7$ superconformal index}\label{app:IE7}
Here we give an explicit expression for the superconformal index of the $E_7$ theory with the flavor fugacities turned off:
\begin{align*}
\mathcal{I}_{E_7}&=1+\frac{133 \ft^4}{v}-134 \ft^6+\frac{134 \ft^7 \chi _2^y}{v}+\ft^8 \left(\frac{7371}{v^2}+v+v^4\right)+\ft^9 \left(-135-v^3\right) \chi _2^y\\&~~+\ft^{10}
\left(-\frac{16148}{v}+v^2+\frac{134 \chi _3^y}{v}\right)+\ft^{11} \left(\frac{16149}{v^2}+v+v^4\right) \chi _2^y\\&~~+\ft^{12} \left(8910+\frac{238602}{v^3}-v^3+\left(-135-v^3\right) \chi _3^y\right)+\ft^{13} \left(\left(-\frac{34103}{v}+v^2\right) \chi _2^y+\frac{134 \chi _4^y}{v}\right)\\&~~+\ft^{14} \left(-\frac{810502}{v^2}-134
v-v^4+v^7+\left(\frac{25193}{v^2}+v+v^4\right) \chi _3^y\right)\\&~~+\ft^{15}
\left(\left(18222+\frac{819413}{v^3}+133 v^3+v^6\right) \chi _2^y+\left(-135-v^3\right) \chi _4^y\right)\\&~~+\ft^{16} \left(\frac{5248750}{v^4}+\frac{944354}{v}-133 v^2+59 v^5+v^8+\left(-\frac{52192}{v}-132 v^2-v^5\right) \chi _3^y+\frac{134
	\chi _5^y}{v}\right)\\&~~+\ft^{17} \left(\left(-\frac{2744244}{v^2}-136 v-132 v^4-v^7\right) \chi_2^y+\left(\frac{34104}{v^2}+v+v^4\right) \chi _4^y\right)\\&~~+\ft^{18} \Bigg(-381096-\frac{23571439}{v^3}+1685 v^3-58 v^6+v^9+\left(27401+\frac{1799756}{v^3}+335 v^3+v^6\right) \chi
_3^y\\&\hspace{350pt}+\left(-135-v^3\right) \chi _5^y\Bigg)\\&~~+\ft^{19} \Bigg(\left(\frac{24534016}{v^4}+\frac{3097067}{v}-3719 v^2-190 v^5+v^8\right) \chi _2^y+\left(-\frac{70282}{v}-266 v^2+v^5\Bigg) \chi
_4^y+\frac{134 \chi _6^y}{v}\right)
\\&~~+ \mathcal{O}(\ft^{20})
\end{align*}

\bibliographystyle{jhep}
\bibliography{ADN1}

\end{document}